\documentclass[12pt]{article}
\usepackage{epsfig}
\usepackage{slashed}

\newcommand{\hl}{\hline}
\newcommand{\ben}{\begin{enumerate}}
\newcommand{\een}{\end{enumerate}}
\newcommand{\bit}{\begin{itemize}}
\newcommand{\eit}{\end{itemize}}
\newcommand{\bc}{\begin{center}}
\newcommand{\ec}{\end{center}}
\newcommand{\bq}{\begin{equation}}
\newcommand{\eq}{\end{equation}}
\newcommand{\bqa}{\begin{eqnarray}}
\newcommand{\eqa}{\end{eqnarray}}

\newcommand{\nn}{\nonumber}
\newcommand{\mtt}[1]{\mbox{\tt{#1}}}
\newcommand{\plaat}[3]{\raisebox{#3pt}{\epsfig{figure=#1.pdf,
width=#2cm}}}

\def\jpsi{{J/\psi}}

\def\ss{{\bigl.^3\hspace{-1mm}S^{[1]}_1}}
\def\sps{{\bigl.^1\hspace{-1mm}S^{[8]}_0}}
\def\so{{\bigl.^3\hspace{-1mm}S^{[8]}_1}}

\def\pj{{\bigl.^3\hspace{-1mm}P^{[8]}_J}}
\def\p0{{\bigl.^3\hspace{-1mm}P^{[8]}_0}}

\def\tsos{{\bigl.^3\hspace{-1mm}S^{[1]}_1}}
\def\oszo{{\bigl.^1\hspace{-1mm}S^{[8]}_0}}
\def\oszs{{\bigl.^1\hspace{-1mm}S^{[1]}_0}}
\def\tsoo{{\bigl.^3\hspace{-1mm}S^{[8]}_1}}
\def\tpjs{{\bigl.^3\hspace{-1mm}P^{[1]}_J}}
\def\tpjo{{\bigl.^3\hspace{-1mm}P^{[8]}_J}}
\def\opos{{\bigl.^1\hspace{-1mm}P^{[1]}_1}}
\def\opoo{{\bigl.^1\hspace{-1mm}P^{[8]}_1}}

\def\oszo{{\bigl.^1\hspace{-1mm}S^{[8]}_0}}
\def\tsoo{{\bigl.^3\hspace{-1mm}S^{[8]}_1}}

\def\to{\rightarrow}

\begin{document}

\pagestyle{empty}


\vspace*{2cm}

\bc\begin{LARGE} {\bf HELAC-Onia: an automatic matrix element generator for heavy quarkonium physics}\\
\end{LARGE} \vspace*{2cm}

{\large {\bf Hua-Sheng Shao}
} \\[12pt]
 {~Department of Physics and State Key Laboratory of Nuclear Physics
and Technology, Peking University,
 Beijing 100871, China}\\{
 ~PH Department,TH Unit,CERN,CH-1211 Geneva 23,Switzerland}
 \\{E-mail:erdissshaw@gmail.com}

\vspace*{1cm}

{\bf ABSTRACT}\\[12pt]  \ec
\begin{quote}

By the virtues of the Dyson-Schwinger equations, we upgrade the
published code \mtt{HELAC} to be capable to calculate the heavy
quarkonium helicity amplitudes in the framework of NRQCD
factorization, which we dub \mtt{HELAC-Onia}. We rewrote the
original \mtt{HELAC} to make the new program be able to calculate
helicity amplitudes of multi P-wave quarkonium states production at
hadron colliders and electron-positron colliders by including new
P-wave off-shell currents. Therefore, besides the high efficiencies
in computation of multi-leg processes within the Standard Model,
\mtt{HELAC-Onia} is also sufficiently numerical stable in dealing
with P-wave quarkonia (e.g. $h_{c,b},\chi_{c,b}$ ) and P-wave
color-octet intermediate states. To the best of our knowledge, it is
a first general-purpose automatic quarkonium matrix elements
generator based on recursion relations on the market.

\end{quote}

%
%



\newpage
\pagestyle{plain}


\bc {\bf PROGRAM SUMMARY}\\[18pt]\ec
{\it Program title:} \\
{\tt HELAC-Onia}. \\[8pt]
{\it Catalogue number:}  \\[8pt]
{\it Program obtainable from:}\\ http://helac-phegas.web.cern.ch/helac-phegas\\
[8pt]
{\it Licensing provisions:} none\\[8pt]
{\it Operating system under which the program has been tested:}\\
Windows, Unix.\\[8pt]
{\it Programming language:} \\
FORTRAN 90\\[8pt]
\\[8pt]
{\it Nature of physical problem:}\\
An important way to explore the law of the nature is to investigate
the heavy quarkonium physics at B factories and hadron colliders.
However, its production mechanism is still unclear, though NRQCD can
explain its decay mechanism in a sufficiently satisfactory manner.
The substantial K-factor in heavy quarkonium production processes
also implies that the associated production of quarkonium and a
relatively large number of particles may paly a crucial role in
unveiling its production mechanism.
\\[8pt]
{\it Method of solution:} \\
A labor-saved and efficient way is to make the tedious amplitudes
calculation automatic. Based on a recursive algorithm derived from
the Dyson-Schwinger equations, the goal of automatic calculation of
heavy quarkonium helicity amplitudes in NRQCD has been achieved.
Inheriting from the virtues of the recursion relations with the
lower computational cost compared to the traditional Feynman-diagram
based method, the multi-leg processes ( with or without
multi-quarkonia up to P-wave states) at colliders are also
accessible.
\\[8pt]
{\it CPC classification code:}
4.4 Feynman Diagrams,11.1 General, High Energy Physics and Computing,11.2 Phase Space and Event Simulation,11.5 Quantum Chromodynamics, Lattice Gauge Theory
\\[8pt]
{\it Typical running time:}
It depends on the process that is to be calculated. However, typically, for all of the tested processes, they take from several minutes to tens of minutes.
\newpage




\section{Introduction}
Studies of heavy-quarkonium systems,e.g. $J/\psi,\Upsilon~\rm{and}
~B_c$,provides an important opportunity to investigate quantum
chromodynamics (QCD) at hadronic level with the least artificial
non-perturbative input parameters by hands. The fact relies on the
non-relativistic property formed by relatively heavy charm and
bottom quarks. Theoretically, these meson can be described well by
non-relativistic QCD (NRQCD)\cite{Bodwin:1994jh} effective theory
with only price that some non-perturbative parameters should be
determined in prior. Although, in principle, the number of these
parameters are not finite , for majority concerned phenomenology
analysis, the number of important parameters are always limited
given in velocity scaling rule. They can be determined once for all
due to their universality property in the effective theory. In fact,
NRQCD was established on a factorization theorem that a perturbative
high-energy exchange part and a process-independent low-energy part
works well in the production and decay processes of heavy
quarkonium. The factorization has been proven rigorously for
quarkonium decay to all orders, while the theorem in production
processes is still absence of proof beyond two-loops. On the
phenomenology side, the inconsistency of NRQCD factorization, cross
section and polarization of heavy quarkonium production has not been
eliminated yet\cite{Bodwin:2012ft}. Hence,fair to say, the mechanism
of heavy quarkonium production is still unclear by now.

The motivation of the paper is to develop an automatic tool for
performing investigations in heavy quarkonium physics.Compared to
some of the traditional Feynman-diagram based
tools\cite{Artoisenet:2007qm,Wang:2004du},the package inherits the
abilities of
\mtt{HELAC}\cite{Kanaki:2000ey,Papadopoulos:2006mh,Cafarella:2007pc},which
is based on a recursive algorithm and hence reduces the
computational cost that grows asymptotically as $n!$ to $a^n$ with
$a \sim 3$, where $n$ is the number of external legs in the
considered process. Therefore, we expect it provides a more
efficient way for people to do physical analysis with multi-particle
processes, which might be especially important in quarkonium
physics\cite{Artoisenet:2008fc}. 

Now we are intending to give a short list of the features in our program compared with {\mtt MADONIA}\cite{Artoisenet:2007qm}.
Since {\mtt HELAC-Onia} is calculated with the recursion relations, the speed of matrix element calculation is expected to be higher than that by {\mtt MADONIA}. 
We have tested a simple process $gg\to c\bar{c}\jpsi(\ss)$. The time in generating one unweighted event by {\mtt HELAC-Onia} is less than that by {\mtt MADONIA}  about a factor of $4$.
Furthermore, {\mtt HELAC-Onia} is quite suitable to calculate multi-quarkonia production, while in {\mtt MADONIA} the number of quarkonium  is restricted to be one. In {\mtt MADONIA}, the computation
 of P-wave amplitude is performing by a numerical derivation, which is expected to be numerical unstable in multi P-wave quarkonia production, while in {\mtt HELAC-Onia} this problem is cured by introducing new
P-wave currents which will be described in the following sections. Another advantage in {\mtt HELAC-Onia} is that it is much easier in its use. On the other hand, because {\mtt MADONIA} is based on Feynman diagrams, it is much more flexible to select some specific diagrams like some diagrams via some specific s-channel propagators, which is difficult to realize in recursion relations. While in {\mtt HELAC-Onia} the feasible processes are only restricted to $pp(\bar{p})$ and $e^-e^+$ collisions at least now,  {\mtt MADONIA} is able to be applied at more colliders as well as subsequent decays.

Before closing this section, we describe the organization of the paper. In
section 2, the recursive algorithm in \mtt{HELAC} is revised.In
section 3, we demonstrate the strategies in helicity amplitudes
calculations of heavy quarkonium production in \mtt{HELAC-Onia}, and
especially the description of P-wave off-shell currents. Several
benchmark processes are computed in section 4. Finally, we explain
our program and draw our conclusions and outlooks in the last two
sections respectively.


%


\section{The recursive algorithm}

The algorithm of
\mtt{HELAC}\cite{Kanaki:2000ey,Papadopoulos:2006mh,Cafarella:2007pc}
is based on the Dyson-Schwinger
equations\cite{Dyson:1949ha,Schwinger:1951ex,Schwinger:1951hq},
which is an generalized version of the Berends-Giele off-shell
recursive relation\cite{Berends:1987me}. To illustrate it, we
consider a process with $n$ external legs. The momenta of these
external legs are denoted as $p_1,p_2,\ldots,p_n$, and the quantum
numbers (e.g. color, helicity) are defined by
$\alpha_1,\alpha_2,\ldots,\alpha_n$. An off-shell current with $k$
external legs can be represented as
\bq\mathcal{J}(\{p_{i_1},\ldots,p_{i_k}\};\{\alpha_{i_1},\ldots,\alpha_{i_k}\})\equiv\plaat{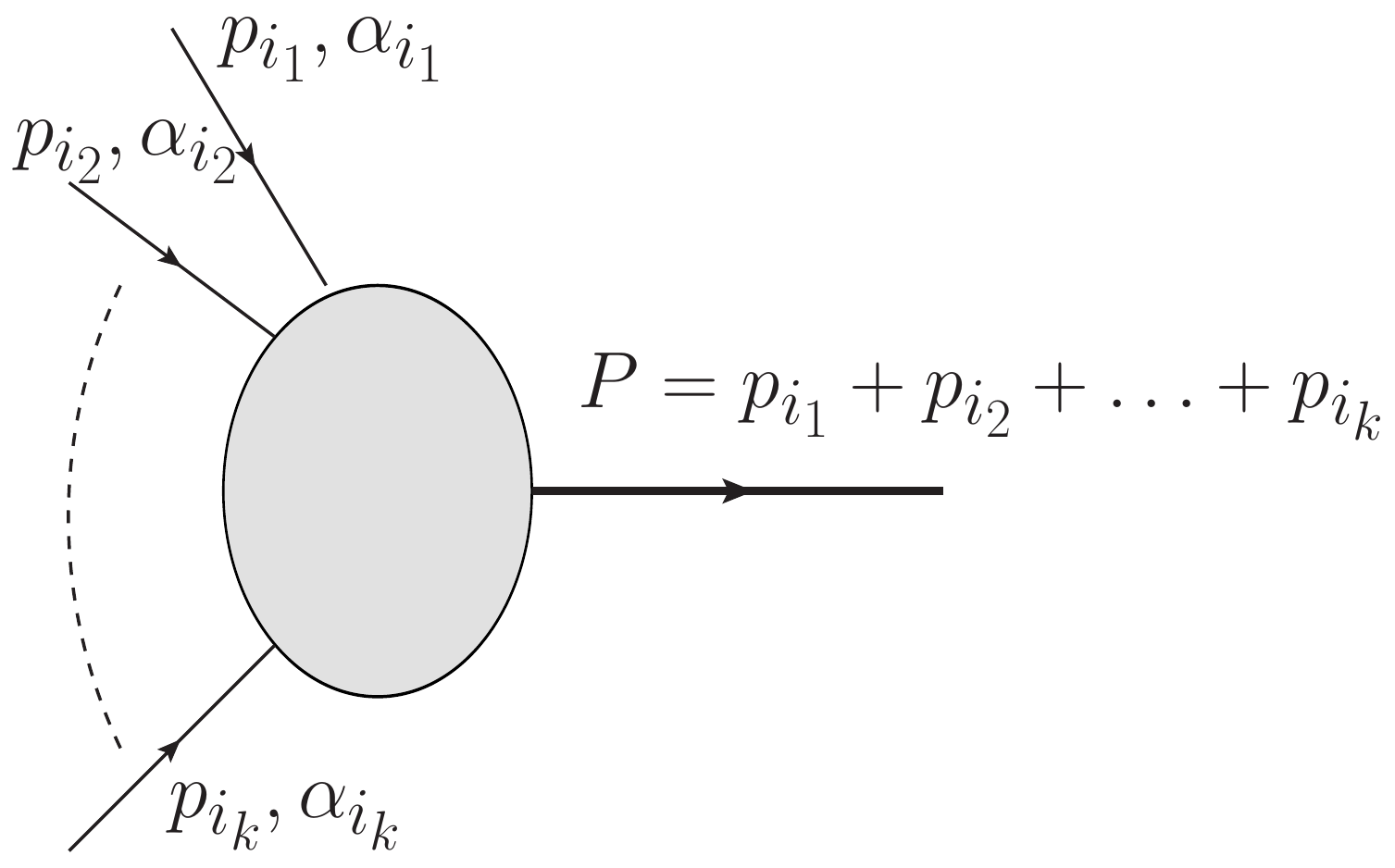}{6}{-45}.\eq
All of the subgraphs that are able to transfer the $k$ external legs
into the off-shell current $\mathcal{J}$ according to the Feynman
rules in the considered model have been included into the shade
bubble. Every current is assigned by a "level" $l$, which is defined
as the number of external legs involved in the current, i.e. the
"level" of
$\mathcal{J}(\{p_{i_1},\ldots,p_{i_k}\};\{\alpha_{i_1},\ldots,\alpha_{i_k}\})$
is $k$. In special, the "level" of each external leg is 1. Then,in
general, all of the currents with higher "level" can be constructed
from those with lower "level". The starting point of the recursion
relation is from the $l=1$ currents, i.e. the external legs. For the
$i$-th external leg, the corresponding current\footnote{The "level"
1 current is on-shell instead of off-shell.} is its
wavefunction\bq\mathcal{J}(\{p_i\};\{\alpha_i\})\equiv\plaat{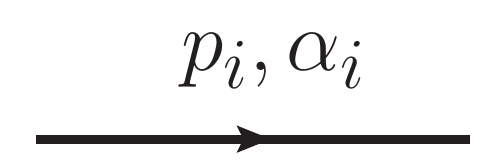}{3}{-4}.\eq
Specifically,for a vector boson
\bq\mathcal{J}(\{p_i\};\{\mu,\lambda\})\equiv\epsilon^{\mu}_{\lambda}(p_i),\eq
where $\mu$ is the lorentz index and $\lambda$ is the helicity of
the vector boson ($\lambda=\pm1$ for a massless vector, while
$\lambda=\pm1,0$ for a massive vector)\footnote{Note that, for
simplicity,we have suppressed other possible quantum numbers like
color for gluon.}, while for a spin-$\frac{1}{2}$ fermion
\bqa\mathcal{J}(\{p_i\};\{+1,\lambda\})\equiv\left\{
\begin{array}{ll}
u_\lambda(p_i)
&\mbox{when $p_i^0\geq0$} \\
v_\lambda(-p_i)
&\mbox{when $p_i^0\leq0$} \\
\end{array}
\right., \nn\\
\mathcal{J}(\{p_i\};\{-1,\lambda\})\equiv\left\{
\begin{array}{ll}
\bar{u}_\lambda(p_i)
&\mbox{when $p_i^0\geq0$} \\
\bar{v}_\lambda(-p_i)
&\mbox{when $p_i^0\leq0$} \\
\end{array}
\right.,\eqa where $+1$ and $-1$ means fermion flow and anti-fermion
flow respectively, $\lambda$ is the helicity index with
$\lambda=\pm1$. The explicit expressions of these $l=1$
wavefunctions are presented in the appendix of
Ref.\cite{Kanaki:2000ey}. The currents with $l=k>1$ can be
constructed from the currents with lower $l$\footnote{For simplicity, we
only consider tri-linear and quadri-linear couplings. However, it is
straightforward to include higher-point vertices as well.}\bq
\plaat{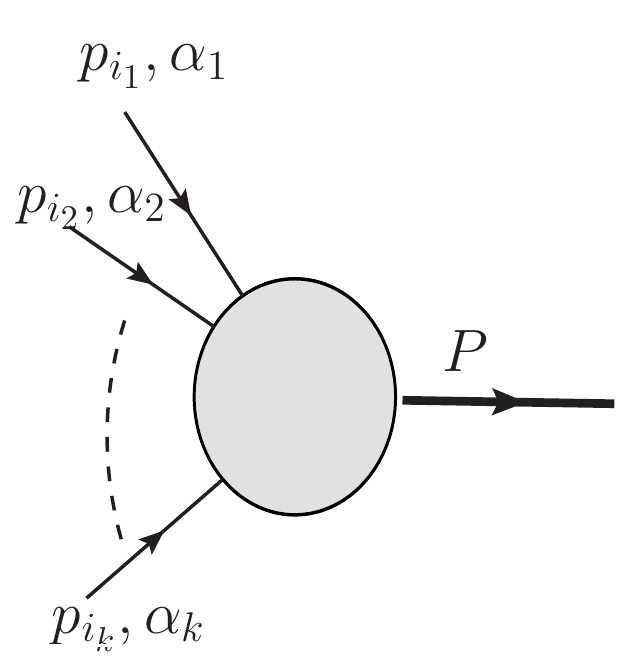}{4}{-40}=\sum_{\sigma}{\sum^{r+s=k}_{r,s>0}{\plaat{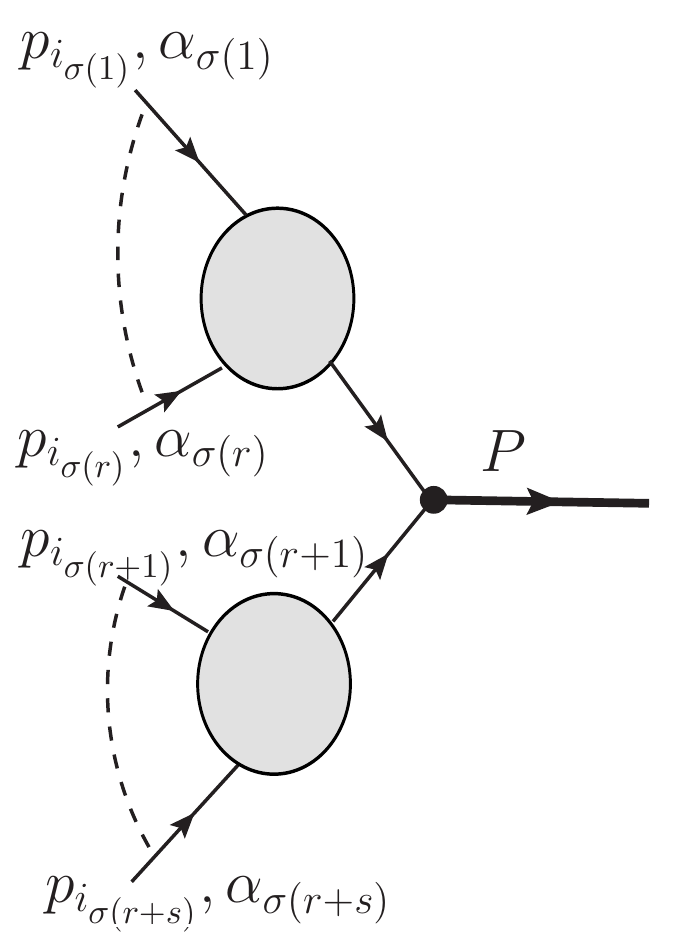}{3.5}{-55}}}\;
+\sum_{\sigma}{\sum^{r+s+t=k}_{r,s,t>0}{\plaat{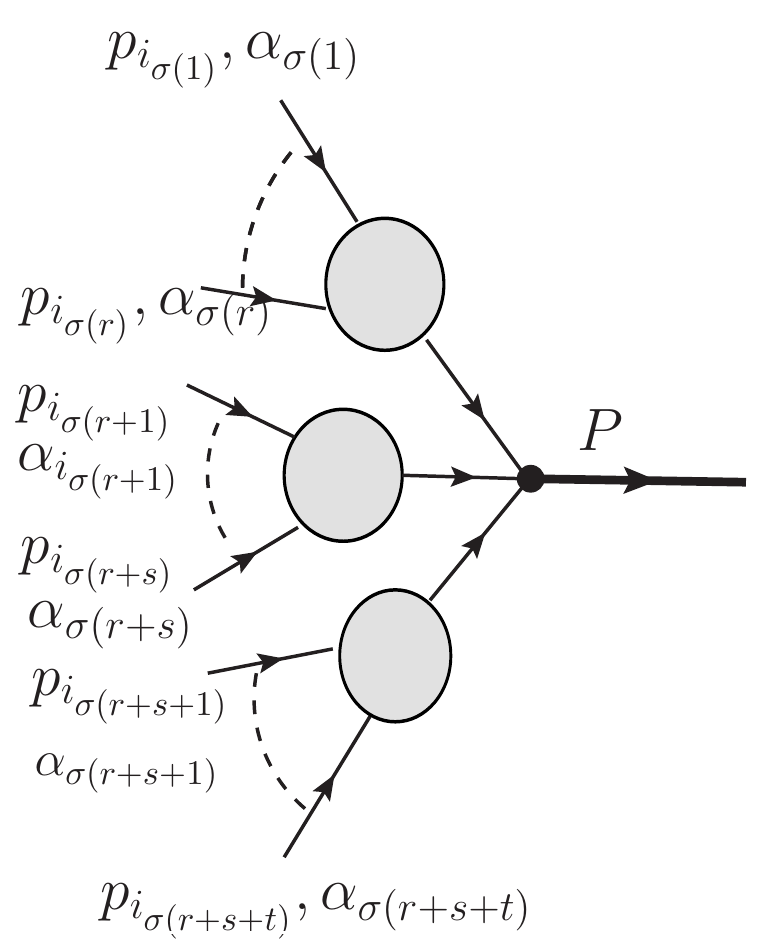}{4}{-60}}},\eq
where $\sigma$ means to exhaust all possible generating "level"
$r,s$ (and $t$) currents formed by the $i_1,\ldots,i_k$ external
legs.Each off-shell currents should be multiplied by its propagator.
The end of the recursion is the forming of the "level" $n$ current,
in which we choose all $l=n-1$ currents\footnote{Actually, they are
on-shell instead of off-shell.} to multiply with the first external
particle's wavefunction. If the flavor of the first external
particle is not exactly same as the flavor of the $l=n-1$ current,
the current is dropped. Finally, we obtain the resulting amplitude.
In this way, one can avoid computing identical sub-graphs
contributing different Feynman diagrams more than once. The
summation of the all sub-graphs that contribute to a specific
current also reduces the number of objects that should be used in
the next "level" recursion formula. Therefore, the computation
complexity will be reduced from $\sim n!$ in Feynman-diagram based
algorithm to $\sim a^n$ in the Dyson-Schwinger based recursive
algorithm, where $a\sim 3$.

In the original \mtt{HELAC} program\cite{Kanaki:2000ey},it uses a
binary representation of the momenta involved in the considered
process\cite{Caravaglios:1995cd}. For each of the external momenta
$p_1,\ldots,p_n$, its binary representation is $2^{i-1}$ for $i$-th
external leg with momenta $p_i$, while for a $l=k$ current
$\mathcal{J}(\{p_{i_1},\ldots,p_{i_k}\};\{\alpha_{i_1},\ldots,\alpha_{i_k}\})$,
it is expressed as $\sum^k_{j=1}{2^{i_j-1}}$. Then, each momentum
$P^{\mu}=\sum^n_{j=1}{m_j~p_{j}}$ can be uniquely expressed by an
integer $m=\sum_{j=1}^n{2^{j-1}m_j}$ where $m_j=0$ or $1$. In this
case, the "level" of an current with momentum
$P^{\mu}=\sum^n_{j=1}{m_j~p_{j}}$ can be calculated directly by
$l=\sum^n_{j=1}{m_j}$. In this case, the sign factor from the
anti-symmetric property of fermions is obtained by
\bq\epsilon(P_1,P_2)=(-1)^{\chi(P_1,P_2)},\chi(P_1,P_2)=\sum^2_{i=n}{\hat{m}_{1i}\sum^{i-1}_{j=1}{\hat{m}_{2j}}}\eq,
with \bqa
P_1&=&\sum^n_{j=1}{m_{1j}p_j},\nonumber\\P_2&=&\sum^n_{j=1}{m_{2j}p_j},\nonumber\\
\hat{m}_{1j}&=&\left\{\begin{array}{ll} 0
&\mbox{when particle $j$ is a boson} \\
m_{1j} &\mbox{when particle $j$ is a fermion}
\\\end{array},\right.\nonumber\\\hat{m}_{2j}&=&\left\{\begin{array}{ll} 0
&\mbox{when particle $j$ is a boson} \\
m_{2j} &\mbox{when particle $j$ is a fermion}
\\\end{array}.\right.\eqa
If the current is constructed by a tri-linear coupling with the
lower "level" currents $P_1$ and $P_2$, it should be multiplied by a
factor $\epsilon(P_1,P_2)$. If it is constructed by a quadri-linear
coupling with currents $P_1$,$P_2$ and $P_3$, the sign factor is
$\epsilon(P_1,P_2,P_3)=\epsilon(P_1,P_2)\epsilon(P_1+P_2,P_3)$.

The way of the color treatment is also an interesting topic in the
matrix element generator. In \mtt{HELAC}, it is using the widely
used color flow basis, which was first proposed in
Ref.\cite{'tHooft:1973jz} and was applied in perturbative QCD
computations in Refs.\cite{Kanaki:2000ms,Maltoni:2002mq}. Basically,
a color octet gluon field $A^a_{\mu}$ is replaced by a $3\times3$
matrix
$(\mathcal{A}_{\mu})^i_j=\frac{1}{\sqrt{2}}A^a_{\mu}(\lambda^a)^i_j$,
where $\lambda^a$ is the Gell-Mann matrix,i.e.
$\textbf{8}=\textbf{3}\otimes\bar{\textbf{3}}-\textbf{1}$, while
incoming quarks or outgoing antiquarks still maintain in the
$\textbf{3}$ representation of SU(3) and outgoing quarks or incoming
antiquarks are in $\bar{\textbf{3}}$ representation. After this
substitution, only the Kronecker notation $\delta$s appear in the
Feynman rules. All the Feynman rules in the color-flow basis have
been established in Ref.\cite{Maltoni:2002mq}. If there are $n_g$
external gluons (denote as $1,2,\ldots,n_g$) and $n_q$ external
quark-antiquark pairs (denote as $n_g+1,n_g+2,\ldots,n_g+n_q$) in
the considered process, the color basis for the amplitude will be in
the form of
\bq\mathcal{C}_i=\delta^1_{\sigma_i(1)}\ldots\delta^{n_g+n_q}_{\sigma_i(n_g+n_q)},\eq
where $\sigma_i$ represents the i-th permutation of
$1,2,\ldots,n_g+n_q$. There are totally $(n_g+n_q)!$ color basis
,though some of them will vanish. With this basis, one can construct
the color matrix via\bq
M_{ij}=\sum{\mathcal{C}_i\mathcal{C}^*_j},\eq and obtain the final
square of matrix elements by
\bq|\mathcal{M}|^2=\sum^{(n_g+n_q)!}_{i,j=1}{A_iM_{ij}A^*_j},\eq
where $A_i$ and $A_j$ are the color-stripped amplitudes.

In order to improve the computation efficiency, a Monte Carlo
sampling over the helicity configurations is adopted in the
program\cite{Kanaki:2000ey} to perform the helicity summation. The
basic idea of this technology is simple. Let us take a massive vector
boson for example. A massive vector boson has three helicity states
$\lambda=\pm1,0$ with wavefunctions
$\epsilon^{\mu}_{+},\epsilon^{\mu}_{-},\epsilon^{\mu}_{0}$. The
strategy puts the concrete helicity summation of
$\sum_{\lambda=\pm,0}{\epsilon^{\mu}_{\lambda}(\epsilon^{\nu}_{\lambda})^*}$
into a continue integration by defining
$\epsilon^{\mu}_{\phi}\equiv\sum_{\lambda=\pm,0}{e^{i\lambda\phi}\epsilon^{\mu}}$.
Then, the summation becomes
$\int^{2\pi}_{0}{\rm{d}\phi\epsilon^{\mu}_{\phi}(\epsilon^{\nu}_{\phi})^*}$,
which can be calculated easily by a Monte Carlo program.

\section{Quarkonium amplitudes in NRQCD}
In the framework of the NRQCD factorization, the cross section of a
heavy quarkonium production can be written as a combination of the
perturbative short-distance parts and the non-perturbative
long-distance matrix elements. For example at the proton-proton
collider, the factorized form of a heavy quarkonium $\mathcal{Q}$
production is written
as\bq\sigma(pp\to\mathcal{Q}+X)=\sum_{i,j,n}{\int{\rm{d}x_1\rm{d}x_2\textit{f}_{i/p}(x_1)\textit{f}_{j/p}(x_2)
\hat{\sigma}(ij\to
Q\bar{Q}[n]+X)\langle\mathcal{O}^{\mathcal{Q}}_n\rangle}},\eq where
$\textit{f}_{i/p}$ and $\textit{f}_{j/p}$ are the parton
distribution functions,$\hat{\sigma}(ij\to Q\bar{Q}[n]+X)$ is the
short distance cross section of producing a heavy quark pair
$Q\bar{Q}$ in a specific quantum state $n$, and
$\langle\mathcal{O}^{\mathcal{Q}}_n\rangle$ represents as the long
distance matrix element. In principle,  for a specific quarkonium
$\mathcal{Q}$,there are infinity number of Fock states $n$ and
infinity number of long distance matrix elements
$\langle\mathcal{O}^{\mathcal{Q}}_n\rangle$. The power counting
rules in NRQCD tell us for any quarkonium, there are only limited
Fock states should be involved in our calculations up to a specific
order of $v$, where $v$ is the relative velocity of the heavy quark
pair that forms the quarkonium. It makes NRQCD be predictable for
hadrons. For example, in the process of $\jpsi$ production, there
are only four different Fock states (i.e. $\ss,\so,\pj$ and
$\sps$)\footnote{We write the Fock states in the spectroscopic form
of $n={\bigl.^{2S+1}\hspace{-1mm}L^{[c]}_J}$, where $S,L,J$ identify
the spin, orbital momentum, total angular momentum states
respectively, and $c=1,8$ means that the intermediate state
$Q\bar{Q}$ can be in color-singlet or color-octet state.}
contributing to its cross section up to $v^7$. The color-singlet
long distance matrix element can be estimated from the
phenomenological models like potential models, while the color-octet
long distance matrix elements can only be determined from the
experimental data till now.

\subsection{Projection method}

To evaluate the process-dependent short distance coefficients, one
has to constraint the $Q\bar{Q}$ into a specific quantum state. A
convenient way to do it is performing projection.

Specifically, the color projectors to the process $ij\to
Q\bar{Q}[{\bigl.^{2S+1}\hspace{-1mm}L^{[c]}_J}]+X$
are\cite{Petrelli:1997ge} $\frac{\delta_{ij}}{N_c}$ when $c=1$ and
$\sqrt{2}\lambda^a_{ij}$ when $c=8$, where $i,j$ are the color
indices of the heavy quark pair $Q\bar{Q}$ and $\lambda^a$ is the
Gell-Mann matrix. The color octet projector which contains the
Gell-Mann matrix will be decomposed into the color-flow basis in the
\mtt{HELAC-Onia}. Moreover, after projecting, no color indices of
the heavy quark pair in the color-singlet states will appear.

Another important constraint of the heavy quark pair is their total
spin. The spin projectors were first derived in
Refs.\cite{Guberina:1980dc,Berger:1980ni}. The general form of the projectors
is\footnote{In \mtt{HELAC-Onia}, we also generalize the projectors
in the case of the heavy quarks in different flavors that form a
heavy quarkonium like $B_c$. But for simplicity, we only consider
the same flavor case here.}
\bq-\frac{1}{2\sqrt{2}(E+m_{Q})}\bar{v}(p_2,\lambda_2)\Gamma_S\frac{\slashed{P}+2E}{2E}u(p_1,\lambda_1),\label{eq:spinpro}\eq
where $m_{Q}$ is the mass of the heavy quark,$p_1,p_2$ and
$\lambda_1,\lambda_2$ are the heavy quarks' momenta and helicity
respectively, $P^{\mu}=p^{\mu}_1+p^{\mu}_2$ is the total momentum of
the heavy quark pair and $E=\frac{\sqrt{P^2}}{2}$. The $\Gamma_S$ is
$\gamma_5$ when $S=0$, and it is
$\epsilon_{\mu}^{\lambda_s}\gamma^{\mu}$ when $S=1$, where
$\lambda_s=\pm,0$ is the helicity of the quarkonium $\mathcal{Q}$
and $\epsilon_{\mu}^{\lambda_s}$ is the polarization vector for the
spin triplet state. For S-wave and P-wave states without
relativistic corrections, $E$ can be set as $m_Q$ directly. After
applying the spin projection, the two external wavefunctions for
open $Q$ and $\bar{Q}$ will be glued. It results in a problem in the
recursive relation, because the recursion begins from the external
wavefunctions. In order to cure it, we decide to cut the glued
fermion chain at the place of $\slashed{P}+2E$ in the projector
Eq.(\ref{eq:spinpro}). Using the completeness relation of
$\slashed{P}+2E=\sum_{\lambda^{\prime}=\pm}{u(P,\lambda^{\prime})\bar{u}(P,\lambda^{\prime})}$,
we use the new "wavefunction" for $Q$ as
$\frac{1}{m_Q}\bar{u}(P,\lambda^{\prime})(\slashed{p}_1+m_Q)$ and
for $\bar{Q}$ as
$-\frac{1}{8\sqrt{2}m_Q}(\slashed{p}_2-m_Q)u(P,\lambda^{\prime})$.
Considering the $\lambda^{\prime}$ in the "wavefunctions" of $Q$ and
$\bar{Q}$ should be exactly same, we have to perform the direct
summation of $\lambda^{\prime}$ in stead of Monte Carlo sampling in
the \mtt{HELAC-Onia}.

\subsection{P-wave currents in \mtt{HELAC-Onia}}
The P-wave calculations are always necessary in the NRQCD
predictions for both P-wave states $h_{c/b},\chi_{c/b}$ and S-wave
states $J/\psi,\Upsilon,\eta_{c/b}$. For example, the color-octet
P-wave states $\pj$ are playing special roles in $\jpsi$
hadroproduction\cite{Ma:2010yw,Butenschoen:2010rq,Butenschoen:2012px,Chao:2012iv,Gong:2012ug}
and photoproduction\cite{Butenschoen:2009zy,Butenschoen:2011ks}.
Although they are power suppressed in NRQCD compared to $\ss$,
fragmentation topologies make them overwhelming the color-singlet
one at the medium and high transverse momentum regime. Hence,
\mtt{HELAC-Onia} is designed to be able to handle with P-wave states
as well with a numerical stable method by introducing new P-wave
off-shell currents.

After expanding the relative momentum
$q^{\nu}=\frac{p^{\nu}_1-p^{\nu}_2}{2}$ between the heavy quark pair
in the amplitude $\mathcal{A}(ij\to Q(p_1)\bar{Q}(p_2)+X)$ in the
non-relativistic approximation, the formula for the calculation of
P-wave amplitude is
\bq\left.(\epsilon^{\lambda_l}_{\nu})^*\frac{\partial}{\partial
q_{\nu}}\mathcal{A}(ij\to
Q(p_1)\bar{Q}(p_2)+X)\right|_{q=0},\label{eq:pwave}\eq where
$\lambda_l=\pm,0$ is the helicity configuration of the polarization
vector $\epsilon^{\lambda_l}_{\nu}$ for P-wave state. The treatment
of the new "wavefunctions" definition of the heavy quark pair avoids
the derivation of the spinors, which might result in numerical
instability.

Alternatively, one could also do a direct numerical derivation by
keeping the small relative momentum $q$ of the quark and antiquark
that forms the heavy quarkonium and approaching $q$ to zero in the
quarkonium rest frame\cite{Artoisenet:2007qm}. However, this direct
numerical derivation might result in numerical unstable potentially
especially when there are many P-wave states involved in the
process.

In contrast, the P-wave currents, which are extended from the
original off-shell currents at parton level, can be written in a
much compact manner. In the \mtt{HELAC-Onia}, we assign each current
with an derivation index, which is also in binary representation.
Assuming there are $n_P$ P-wave states in the considered process,
the relative momenta of the $i$-th heavy quark pair that forms
P-wave state is denoted as $q_i$ where $i=1,\ldots,n_P$. The general
derivation index form for a current is
$b=\sum^{n_P}_{i=1}{b_i2^{i-1}}$ with $b_i=0$ or $1$. If the current
has been derived by $q_i$ as done like in Eq.(\ref{eq:pwave}),$b_i$
is $1$, otherwise $b_i=0$. Finally , only the amplitudes with
$b=2^{n_P}-1$ are kept. The numerical stable form of P-wave currents
avoids the large numerical cancellation.

\section{Benchmark processes}
We are in the position to illustrate the validation and applications
of \mtt{HELAC-Onia} to the heavy quarkonium production at the
proton-proton, proton-antiproton and electron-positron colliders.
\subsection{$B_c$ meson production at the LHC}
$B_c$ production is an interesting channel to investigate QCD, and
it has been widely studied at the hadron
colliders\cite{Chang:2003cq,Chang:2005hq,Berezhnoy:1997er}. The
available results have been used by the
\mtt{MADONIA}\cite{Artoisenet:2007qm} for testing the correctness of
the code. We will also compare our results calculated by the
\mtt{HELAC-Onia} with those presented in
Ref.\cite{Artoisenet:2007qm}. We only consider the $B_c$ production
at the LHC with the center-of-mass energy $14$ TeV with the initial
gluon-gluon fusion and quark-antiquark annihilation here. All of the
input parameters are taken as same as those in
Ref.\cite{Artoisenet:2007qm}:
\begin{itemize}
\item[(1)]The masses of the bottom, charm quarks and $B_c$ meson are
set as
$m_b=4.9~\rm{GeV},m_c=1.5~\rm{GeV},m_{B_c}=m_b+m_c=6.4~\rm{GeV}$.
\item[(2)]The parton distribution function (PDF) set is chosen as CTEQ6L1~\cite{Pumplin:2002vw}.
\item[(3)] The factorization scale $\mu_F$ of PDF and the renormalization
scale $\mu_R$ are set as
$\mu_F=\mu_R=2(m_b+m_c)=12.8~\rm{GeV}$.Moreover, the strong coupling
constant is fixed as $\alpha_S(\mu_R)=0.189$.
\item[(4)] The color-singlet long distance matrix elements are taken as $\langle\mathcal{O}^{B_c}({{\bigl.^{2S+1}\hspace{-1mm}S^{[1]}_J})}\rangle
=(2J+1)0.736~\rm{GeV}^3,\langle\mathcal{O}^{B_c}({{\bigl.^{2S+1}\hspace{-1mm}P^{[1]}_J})}\rangle
=(2J+1)0.287~\rm{GeV}^5$.
\item[(5)] The color-octet long distance matrix elements are
related with the color-singlet
ones,i.e.$\langle\mathcal{O}^{B_c}({{\bigl.^{2S+1}\hspace{-1mm}S^{[8]}_J})}\rangle
=\langle\mathcal{O}^{B_c}({{\bigl.^{2S+1}\hspace{-1mm}S^{[1]}_J})}\rangle/100$,
$\langle\mathcal{O}^{B_c}({{\bigl.^{2S+1}\hspace{-1mm}P^{[8]}_J})}\rangle
=\langle\mathcal{O}^{B_c}({{\bigl.^{2S+1}\hspace{-1mm}P^{[1]}_J})}\rangle/100$.
\end{itemize}
Our final results (with Monte Carlo statistical errors) are shown in
the second column of Table.\ref{tab:Bc}, where we also listed the
corresponding results (in the third column of Table.\ref{tab:Bc})
presented in Ref.\cite{Artoisenet:2007qm} for the convenience of
comparison. We find that our results are in agreement with those in
Ref.\cite{Artoisenet:2007qm}.
\begin{table}[h]
\begin{center}
\begin{tabular}{|{c}*{3}{|c}|} \hline
process &  {\mtt HELAC-Onia}(nb) &  {\mtt
MADONIA}(nb)  \\
\hline $gg\rightarrow B_c^+(\oszs)b\bar{c}$ &
$39.3994\pm0.0958382$ & $39.4$ \\
$gg\rightarrow B_c^+(\tsos)b\bar{c}$& $98.3109\pm0.287252$ & $98.3$
\\ $gg\rightarrow B_c^+(\opos)b\bar{c}$ &
$5.21131\pm0.0144431$ & $5.20$ \\$gg\rightarrow
B_c^+(\tpjs)b\bar{c}$ & $16.7341\pm0.0589108$ & $16.72$ \\
$gg\rightarrow B_c^+(\oszo)b\bar{c}$ & $0.411671\pm0.00169734$ &
$0.411$ \\$gg\rightarrow B_c^+(\tsoo)b\bar{c}$ &
$1.78657\pm0.00624756$ & $1.79$ \\$gg\rightarrow
B_c^+(\opoo)b\bar{c}$ & $0.11816\pm0.000754526$ & $0.117$
\\$gg\rightarrow B_c^+(\tpjo)b\bar{c}$ &
$0.305862\pm0.0011841$ & $0.3051$ \\
$q\bar{q}\rightarrow B_c^+(\oszs)b\bar{c}$ &
$0.137782\pm0.000896985$ & $0.137$ \\
$q\bar{q}\rightarrow B_c^+(\tsos)b\bar{c}$ & $0.83905\pm0.00524885$
& $0.834$ \\$q\bar{q}\rightarrow B_c^+(\opos)b\bar{c} $ &
$0.0296125\pm0.000154919$ & $0.0295$ \\
$q\bar{q}\rightarrow B_c^+(\tpjs)b\bar{c}$ &
$0.111259\pm0.000839535$ & $0.1105$ \\
$q\bar{q}\rightarrow B_c^+(\oszo)b\bar{c}$ &
$0.00103294\pm4.44716\cdot10^{-6}$ & $0.00103$
\\
$q\bar{q}\rightarrow B_c^+(\tsoo)b\bar{c}$ &
$0.00707624\pm0.0000459292$ & $0.00703$ \\
$q\bar{q}\rightarrow B_c^+(\opoo)b\bar{c}$ &
$0.000253678\pm2.19206\cdot10^{-6}$ & $0.000251$
\\$q\bar{q}\rightarrow B_c^+(\tpjo)b\bar{c}$ &
$0.000826534\pm5.16988\cdot10^{-6}$ & $0.0008207$\\\hline
\end{tabular}
\end{center}
\caption{\label{tab:Bc}Cross sections of inclusive $B_c^+$
production at the LHC with the center-of-mass energy $14$ TeV. The
data in the third column are taken from
Ref.\cite{Artoisenet:2007qm}. In the second column, the Monte Carlo
statistical errors are also given.}
\end{table}
\subsection{Charmonia production at the B factory}
The charmonia production from the electron-positron collisions has
been extensively studied over the past decade. We do not intend to
recall the long story of the theoretical and experimental studies on
this topic here, which was already summarized in
Ref.\cite{Brambilla:2010cs}. We only want to show the application
and validation of the program \mtt{HELAC-Onia} in calculating the
quarkonium observables in the $e^+e^-$ colliders in this section.

The first theoretical results of the inclusive charmonium
association production with $c\bar c$ via single virtual photon
exchanging at the B factory with the center-of-mass energy $10.6$
GeV was presented in Ref.\cite{Liu:2003jj}, which are only
calculated at the leading order in $\alpha_S$ and $v$. Moreover,the
results of $\eta_c$ and $\jpsi$ production with gluons at the B
factory are given in Ref.\cite{Artoisenet:2007qm}. We put the same
input parameters into \mtt{HELAC-Onia}:
\begin{itemize}
\item[(1)] The charm quark's mass $m_c$ is $1.5~\rm{GeV}$, and the
masses of the charmonia considered are $2m_c=3~\rm{GeV}$. The mass
of the electron and positron are safely ignored.
\item[(2)]The renormalization scale $\mu_R$ is chosen as
$2m_c=3~\rm{GeV}$. In this way, the strong coupling constant is
fixed as $\alpha_S(\mu_R)=0.26$, while the electromagnetic fine
structure constant is also set as $\alpha=1/137$.
\item[(3)] The color-singlet long distance matrix elements are $\langle\mathcal{O}({{\bigl.^{2S+1}\hspace{-1mm}S^{[1]}_J})}\rangle
=(2J+1)0.387~\rm{GeV}^3$.
\end{itemize}
Our color-singlet S-wave results are presented in
Table.\ref{tab:ee1}, from which we see that all of our results are
in good agreement with those in
Refs.\cite{Liu:2003jj,Artoisenet:2007qm}. In the first two rows of
Table.\ref{tab:ee3}, we also presented the cross sections of
$\eta_c$ and $\jpsi$ production in association with $c\bar c$ at
$\mathcal{O}(\alpha^2\alpha_S^2+\alpha^3\alpha_S+\alpha^4)$, which
means that we have include both the single photon exchanging channel
and the double photon exchanging channel. To the best of our
knowledge, they are new.

Several years ago, it was reported that there was a large
discrepancy between the theoretical predictions and the experimental
measurements in the exclusive double charmonia production at the B
factory (see review in e.g.Ref.\cite{Brambilla:2010cs}). The large
discrepancy has attracted a lot of studies in this fields. We
recalculated some of the cross sections at the B factory with
$\sqrt{s}=10.6~\rm{GeV}$ in Table.\ref{tab:ee2} with the same input
as those given in Ref.\cite{Braaten:2002fi}. We listed the input
parameters as following:
\begin{itemize}
\item[(1)]The mass of the charm quark is $1.4~\rm{GeV}$, and the
masses of the charmonia are $2m_c$.
\item[(2)] The renormalization scale $\mu_R$ is set as
$\sqrt{s}/2=5.3~\rm{GeV}$, and $\alpha_S(\mu_R)=0.21,\alpha=1/137$.
\item[(3)] The color-singlet long distance matrix elements are $\langle\mathcal{O}({{\bigl.^{2S+1}\hspace{-1mm}S^{[1]}_J})}\rangle
=(2J+1)0.335~\rm{GeV}^3$ and
$\langle\mathcal{O}({{\bigl.^{2S+1}\hspace{-1mm}P^{[1]}_J})}\rangle
=(2J+1)0.053~\rm{GeV}^5$.
\end{itemize}
The results in Table.\ref{tab:ee2} only include
$\mathcal{O}(\alpha^2\alpha_S^2)$ and leading-order in $v$
perturbative calculations. Good agreement is found between
\mtt{HELAC-Onia} and Ref.\cite{Braaten:2002fi}. The
$\mathcal{O}(\alpha^2\alpha_S^2+\alpha^3\alpha_S+\alpha^4)$ cross
sections for $\jpsi\jpsi$ and $\jpsi h_c$ exclusive productions are
shown in the last two rows of Table.\ref{tab:ee3}. These results
agree with those given in Ref.\cite{Bodwin:2002kk}.

\begin{table}
\bc
\begin{tabular}{|{c}*{3}{|c}|} \hline
process & {\mtt HELAC-Onia}(fb) & Refs.\cite{Liu:2003jj,Artoisenet:2007qm}(fb) \\
\hline $e^+e^-\rightarrow\gamma^*\to\eta_c(\oszs)c\bar{c}$ &
$58.7938\pm0.154193$ & $58.7$
\\$e^+e^-\rightarrow\gamma^*\to \eta_c(\oszs)ggg$ &
$3.72893\pm0.0063512$ & $3.72$
\\$e^+e^-\rightarrow\gamma^*\to J/\psi(\tsos)c\bar{c}$ &
$147.864\pm0.305001$ & $148$
\\$e^+e^-\rightarrow\gamma^*\to J/\psi(\tsos)gg$ &
$266.037\pm0.247366$ & $266$
\\\hline
\end{tabular}
\ec \caption{\label{tab:ee1}Cross sections of the inclusive
$\eta_c$ and $\jpsi$ production via single virtual photon exchanging
at the B factory with the center-of-mass energy $10.6~\rm{GeV}$.}
\end{table}

\begin{table}
\bc
\begin{tabular}{|{c}*{3}{|c}|} \hline
process & {\mtt HELAC-Onia}(fb) & Ref.\cite{Braaten:2002fi}(fb) \\
\hline $e^+e^-\rightarrow\gamma^*\to J/\psi(\tsos)\eta_c(\oszs)$ &
$3.78154\pm0.00338108$ & $3.78$
\\$e^+e^-\rightarrow\gamma^*\to h_c(\opos)\eta_c(\oszs)$ &
$0.308533\pm0.000198459$ & $0.308$\\$e^+e^-\rightarrow\gamma^*\to
J/\psi(\tsos)\chi_{cJ}(\tpjs)$ & $3.47635\pm0.00453553$ & $3.47$
 \\$e^+e^-\rightarrow\gamma^*\to
h_c(\opos)\chi_{cJ}(\tpjs)$ & $0.328299\pm0.000392734$ & $0.328$
\\\hline
\end{tabular}
\ec \caption{\label{tab:ee2}Cross sections of the exclusive double
charmonia production via single virtual photon exchanging at the B
factory with the center-of-mass energy $10.6~\rm{GeV}$.}
\end{table}

\begin{table}
\bc
\begin{tabular}{|{c}*{3}{|c}|} \hline
process & {\mtt HELAC-Onia}(fb) & Ref.\cite{Bodwin:2002kk}(fb) \\
\hline $e^+e^-\rightarrow \eta_c(\oszs)c\bar{c}$ &
$61.6802\pm0.0854359$ & $-$\\$e^+e^-\rightarrow
J/\psi(\tsos)c\bar{c}$ & $166.499\pm0.175318$ & $-$
\\$e^+e^-\rightarrow J/\psi(\tsos)J/\psi(\tsos)$ &
$6.64805\pm0.0123474$ & $6.65$\\$e^+e^-\rightarrow
J/\psi(\tsos)h_c(\opos)$ & $0.00606923\pm6.84416\cdot10^{-6}$ &
$0.0061$
\\\hline
\end{tabular}
\ec \caption{\label{tab:ee3}The
$\mathcal{O}(\alpha^2\alpha_S^2+\alpha^3\alpha_S+\alpha^4)$ cross
sections of the charmonia production at the B factory with the
center-of-mass energy $1.96~\rm{GeV}$.}
\end{table}

\subsection{Double quarkonia production at the Tevatron and the LHC}
Double quarkonia production at the hadron colliders is a useful way
to investigate the color-octet mechanism. In this section, we will
compare the results calculated by \mtt{HELAC-Onia} and those in the
literature\cite{Li:2009ug}. The input parameters (same as those in
Ref.\cite{Li:2009ug}) are:
\begin{itemize}
\item[(1)]$m_c=1.5,m_b=4.9$. The mass of a heavy quarkonium is
just approximated as the sum of its constituent heavy quarks'
masses. In other words, if a heavy quarkonium $H$ is composed by
$Q_1\bar{Q_2}$, then $m_H=m_{Q_1}+m_{Q_2}$.
\item[(2)]$\mu_F=\mu_R=\sqrt{m_H^2+p_T^2}$ for the heavy quarkonium
$H$.
\item[(3)]PDF set is CTEQ6L1~\cite{Pumplin:2002vw}. Therefore, the
running of $\alpha_S$ is evaluated by the leading-order formula in
the PDF set.
\item[(4)] The S-wave color-singlet long distance matrix elements
are $\langle\mathcal{O}^{c\bar
c}({{\bigl.^{2S+1}\hspace{-1mm}S^{[1]}_J})}\rangle
=(2J+1)0.389134~\rm{GeV}^3$,$\langle\mathcal{O}^{b\bar
b}({{\bigl.^{2S+1}\hspace{-1mm}S^{[1]}_J})}\rangle
=(2J+1)2.34722~\rm{GeV}^3$ and $\langle\mathcal{O}^{b\bar
c}({{\bigl.^{2S+1}\hspace{-1mm}S^{[1]}_J})}\rangle
=(2J+1)0.720017~\rm{GeV}^3$.
\item[(5)] The pseudorapidity $\eta$ cuts on the final quarkonia are
$|\eta|<0.6$ at the Tevatron and $|\eta|<2.4$ at the LHC.
\end{itemize}
The S-wave color-singlet cross sections are shown in
Table.\ref{tab:doubleTev} (Tevatron with $\sqrt{s}=1.96~\rm{TeV}$)
and in Table.\ref{tab:doubleLHC} (LHC with $\sqrt{s}=14~\rm{TeV}$).

\begin{table}
\bc
\begin{tabular}{|c|c|c|c|c|c|} \hline Final States &
\mtt{HELAC-Onia}(nb) & Ref.\cite{Li:2009ug}(nb) \\\hline $2\eta_c(\oszs)$ & $3.316\cdot10^{-3}\pm3.705\cdot10^{-6}$
& $3.32\cdot10^{-3}$ \\
$2J/\psi(\tsos)$ & $0.05631\pm4.437\cdot10^{-5}$
& $0.0563$ \\
$2\eta_b(\oszs)$ &
$1.866\cdot10^{-5}\pm2.385\cdot10^{-8}$  & $1.87\cdot10^{-5}$\\
$2\Upsilon(\tsos)$ & $1.226\cdot10^{-4}\pm1.489\cdot10^{-7}$  & $1.23\cdot10^{-4}$ \\
$B_c(\oszs)\overline{B}_c(\oszs)$ &
$3.854\cdot10^{-3}\pm9.529\cdot10^{-6}$ &
$3.86\cdot10^{-3}$ \\
$B_c(\oszs)\overline{B}_c(\tsos)$ &
$1.001\cdot10^{-3}\pm2.492\cdot10^{-6}$ &
$1.00\cdot10^{-3}$ \\
$B_c(\tsos)\overline{B}_c(\tsos)$ &
$8.226\cdot10^{-3}\pm9.531\cdot10^{-6}$ & $8.23\cdot10^{-3}$
\\\hline
\end{tabular}
\ec \caption{\label{tab:doubleTev}Cross sections of double
quarkonium production at the Tevatron with the center-of-mass energy
$1.96~\rm{TeV}$.}
\end{table}
\begin{table}
\bc
\begin{tabular}{|c|c|c|c|c|c|} \hline Final States & \mtt{HELAC-Onia}(nb) & Ref.\cite{Li:2009ug}(nb) \\\hline $2\eta_c(\oszs)$
& $2.730\pm0.01710$ & $2.73$\\
$2J/\psi(\tsos)$
& $2.832\pm1.721\cdot10^{-3}$ & $2.83$ \\
$2\eta_b(\oszs)$ &
$7.373\cdot10^{-3}\pm1.802\cdot10^{-5}$ & $7.36\cdot10^{-3}$ \\
$2\Upsilon(\tsos)$  & $0.01514\pm1.184\cdot10^{-5}$ &
$0.0151$ \\
$B_c(\oszs)\overline{B}_c(\oszs)$ & $0.2723\pm1.461\cdot10^{-4}$&
$0.272$\\
$B_c(\oszs)\overline{B}_c(\tsos)$ & $0.08379\pm4.430\cdot10^{-5}$ &
$0.0837$\\
$B_c(\tsos)\overline{B}_c(\tsos)$ & $0.7078\pm3.797\cdot10^{-4}$ &
$0.708$
\\\hline
\end{tabular}
\ec \caption{\label{tab:doubleLHC}Cross sections of double
quarkonium production at the LHC with the center-of-mass energy
$14~\rm{TeV}$.}
\end{table}
\subsection{Hadroproduction of $\jpsi$ and $\Upsilon$ in association with a heavy-quark pair}
\begin{figure}
\includegraphics[width=0.9\textwidth]{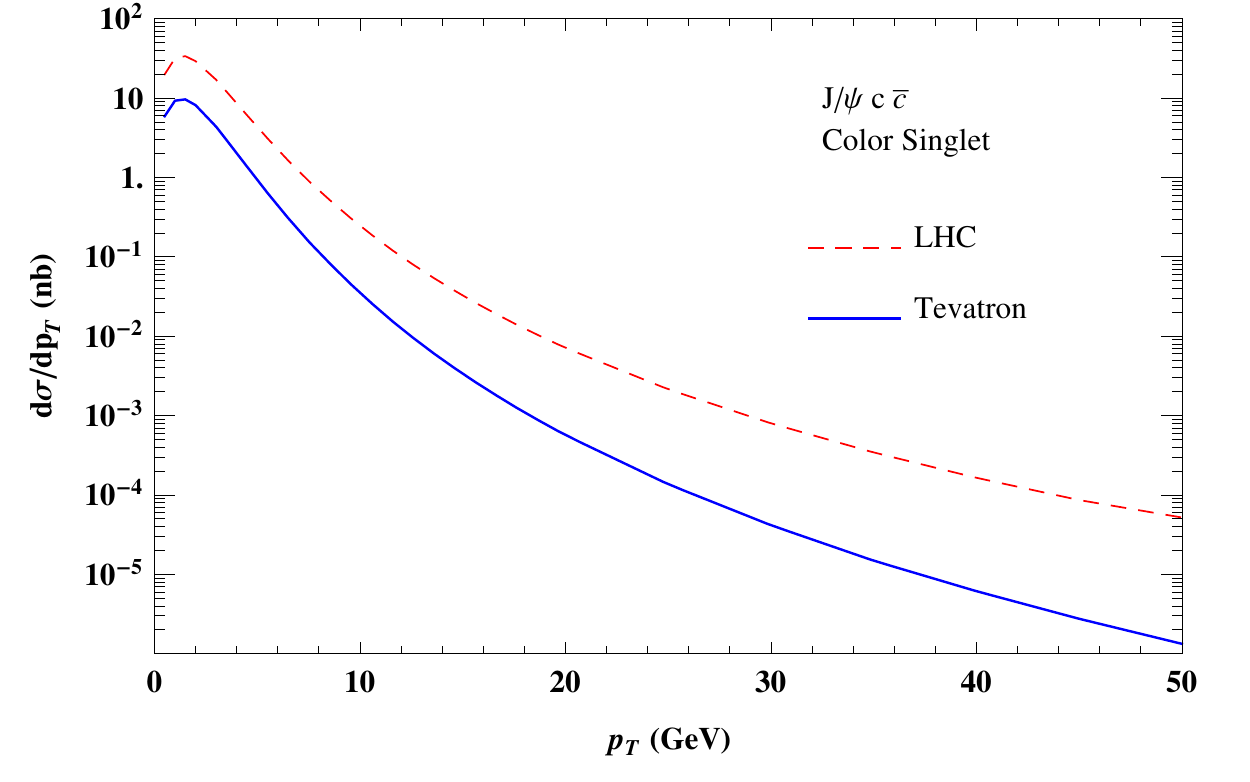}
\includegraphics[width=0.9\textwidth]{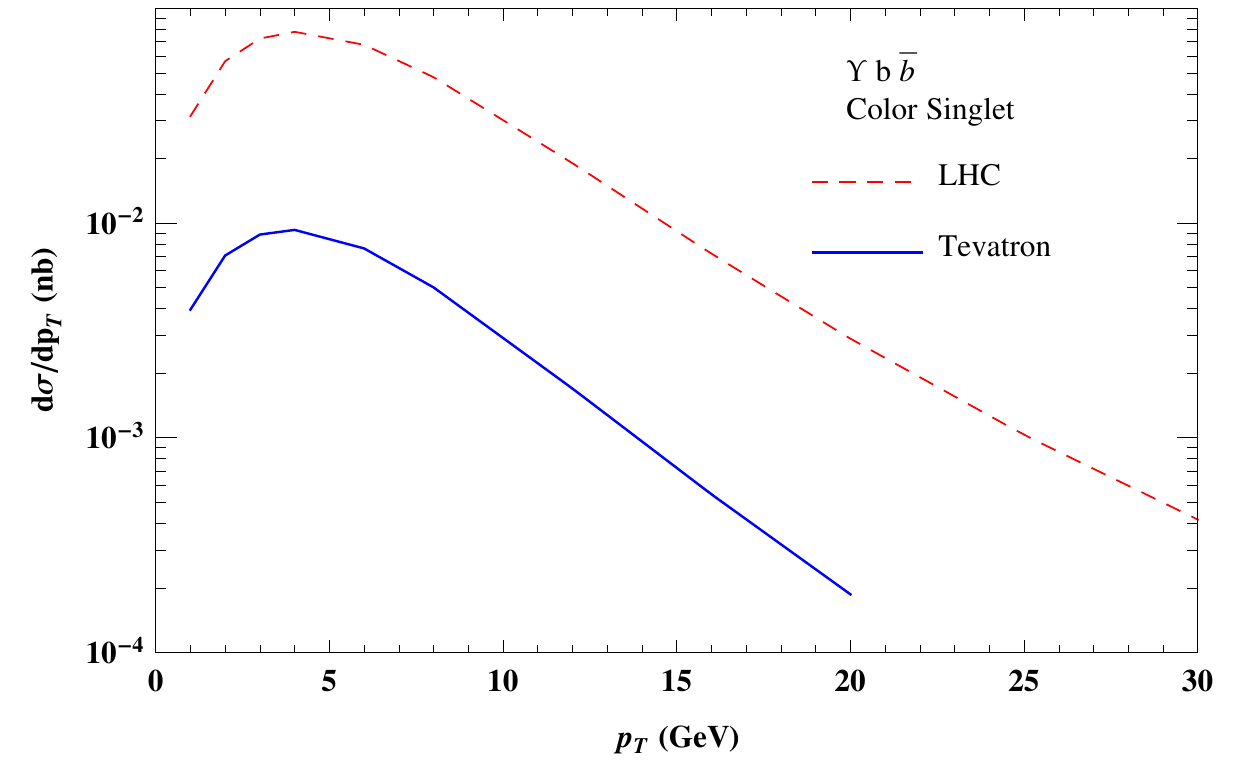}
\caption{\label{fig:jpsiccbar}$p_T$ distributions of
$\jpsi$($\Upsilon$) production in association with $c\bar c$($b \bar
b$) at the Tevatron and LHC. Only color-singlet states are
considered.}
\end{figure}

The measurements of the $\jpsi$ and $\Upsilon$ in association with a
heavy quark pair at the hadron collider are interesting because not
only they will contribute to the inclusive $\jpsi$ and $\Upsilon$
production but also they are useful way to study color-octet
mechanism at the Tevatron and the LHC. In figure
\ref{fig:jpsiccbar}, we present the transverse momentum $p_T$
distributions of $\jpsi$ and $\Upsilon$ via color-singlet channel at
the Tevatron with $\sqrt{s}=1.96~\rm{TeV}$ and the LHC with
$\sqrt{s}=14~\rm{TeV}$. All of our results are in agreement with
those in Ref.\cite{Artoisenet:2007xi}. For completeness, we list our
input for calculation $\jpsi c\bar c$ and $\Upsilon b\bar b$
production as follows:
\begin{itemize}
\item[(1)]$m_c=1.5~\rm{GeV}, m_b=4.75~\rm{GeV}$ and $m_{\jpsi}=2m_c,m_{\Upsilon}=2m_b$.
\item[(2)]$\mu_F=\mu_R=\sqrt{(4m_Q)^2+p_T^2}$, where $m_Q$ is $m_c$ for $\jpsi c\bar c$ and $m_b$ for $\Upsilon b\bar b$.
\item[(3)]PDF set is CTEQ6M~\cite{Pumplin:2002vw}. The running of
$\alpha_S$ is following the next-to-leading order formula in CTEQ.
\item[(4)]The color-singlet long distance matrix elements are $\langle\mathcal{O}^{\jpsi}(\ss)\rangle
=1.16024~\rm{GeV}^3$ and $\langle\mathcal{O}^{\Upsilon}(\ss)\rangle
=9.28192~\rm{GeV}^3$.
\item[(5)] The rapidity cuts are applied as $|y|<0.6$ at the
Tevatron and $|y|<0.5$ at the LHC.
\end{itemize}

\bc
\ec
\subsection{Spin density matrix and polarization}
Besides the total cross sections and other unpolarized observables
like $p_T$ spectrum, \mtt{HELAC-Onia} is also designed to be able to
calculate the spin density matrices of heavy quarkonia. Hence, it
can be taken as a useful tool to calculate the polarization
observables of heavy quarkonia in various polarization frames. It
has been successfully used in:
\begin{itemize}
\item[(1)]the next-to-leading order inclusive $\jpsi$ polarization at
the Tevatron and LHC~\cite{Chao:2012iv};
\item[(2)]the polarization of the inclusive $\chi_c$
hadroproduction~\cite{Shao:2012fs,Chao:2012};
\item[(3)]the polarized $\chi_c$ production in associate with a
charm quark pair at the LHC\cite{Shao:2012}.
\end{itemize}
The readers who are interested in these topics can refer to the
corresponding (forthcoming) publications.


Besides the above examples, there are many other aspects of
quarkonium physics one can perform analysis with
\mtt{HELAC-Onia},for instance, the quarkonium production with jets
(inclusive or exclusive\footnote{Thanks to the useful discussion
with Qiang Li, we have implemented the matrix element and parton
shower matching method MLM
scheme~\cite{Mangano:2006rw,Alwall:2007fs} in the program for quarkonium associated production with jets. The study of this issue in {\mtt HELAC-Onia} will be presented elsewhere.}) and/or
weak bosons. We refrain to illustrate more examples here.

One can feel free to use the modified
\mtt{PHEGAS}\footnote{\mtt{PHEGAS} is modified to generate quarkonium events.}~\cite{Papadopoulos:2000tt,Cafarella:2007pc},\mtt{RAMBO}~\cite{Kleiss:1985gy}
or \mtt{VEGAS}\cite{Lepage:1977sw} to perform Monte Carlo
evaluation. Standard Les Houches Event files~\cite{Boos:2001cv} are
also generated.
\section{Running the program}
We are in the position to describe how can one run the program. The
program is split in two major phases, which were called {\it
initialization phase} and {\it computation phase} in
Ref.\cite{Kanaki:2000ey}.During the {\it initialization phase}, the
program selects all the relevant sub-amplitudes for the required
process and evaluates the color matrix $M_{ij}$, while during the
the {\it computation phase},it computes the amplitude for each phase
space point introduced by \mtt{PHEGAS},\mtt{RAMBO} or \mtt{VEGAS}.

The running of the program is very easy. If the program is running
under the {\tt Unix}, one should specify the {\tt Fortran90}
compilation in the first line of {\tt makefile}. The default one is
{\tt gfortran}. For the {\tt Windows} user, it is running only after
the user has included all of the {\mtt Fortran90} files in his/her
project. There are two input files that should be specified by the
user before running the program. They are {\tt process.inp} and {\tt
user.inp}.

In the {\tt process.inp}, the user should tell the program the
information of the process including the number of external
particles (in the first line) and the ids of the particles (in the
second line). The table of the ids for the particles ( not hadrons )
in the standard-model defined in \mtt{HELAC-Onia} are the same as
that in \mtt{HELAC}, which is shown in Table.\ref{tab:smpart}. The
naming rules of the ids for the heavy quarkonia in \mtt{HELAC-Onia}
are:
\begin{itemize}
\item[(1)] The ids of the heavy quarkonia are in $6$-digits.
\item[(2)] The first two digits are $44$ for charmonia, $55$ for
bottomonia and $45$ for $B_c$.
\item[(3)] The next four digits are just recording the information of which
intermediate Fock states. In general, the four digits are in the
order of $(2S+1)LJc$ for ${\bigl.^{2S+1}\hspace{-1mm}L^{[c]}_J}$.
For example, $3118$ means the intermediate state is
${\bigl.^{3}\hspace{-1mm}P^{[8]}_1}$.
\item[(4)] Charmonia and bottomonia are all self-conjugated mesons, while
$B_c$ are not. A minus sign is used to represent the anti-particle.
In the program, we treat $B_c^+$ as the particle while $B_c^-$ as
the anti-particle. For example, the id of
$B_c^-({\bigl.^{3}\hspace{-1mm}P^{[8]}_1})$ is $-453118$.
\end{itemize}
Using these rules, one can calculate the helicity amplitudes for
S-wave and P-wave quarkonia production from $pp$,$p\bar p$ and
$e^+e^-$ collisions.We take one example. If the user want to
calculate $gg\to c\bar{c}[{\bigl.^{3}\hspace{-1mm}P^{[8]}_1}]+c\bar
c$. The first line of {\tt process.inp} is $5$, and the second line
of {\tt process.inp} is \begin{eqnarray*}
35~35~443118~7~-7.\end{eqnarray*}

\begin{table}
\bc\begin{tabular}{|c|c|} \hl
$\nu_e,e^-,u,d,\nu_\mu,\mu^-,c,s,\ldots$ & $1,\ldots,12$
\\
$\bar{\nu_e},e^+,\bar{u},\bar{d},\bar{\nu}_\mu,\mu^+,
\bar{c},\bar{s},\ldots$ & $-1,\ldots,-12$ \\
$\gamma,Z,W^+,W^-,g$ & $31,\ldots,35$ \\
$H,\chi,\phi^+,\phi^-$ & $41,\ldots,44$ \\
\hl
\end{tabular}\ec
\caption{\label{tab:smpart}The ids of the "elementary" particles in
the standard model in \mtt{HELAC-Onia}.}
\end{table}

\begin{figure}
\includegraphics[width=\textwidth]{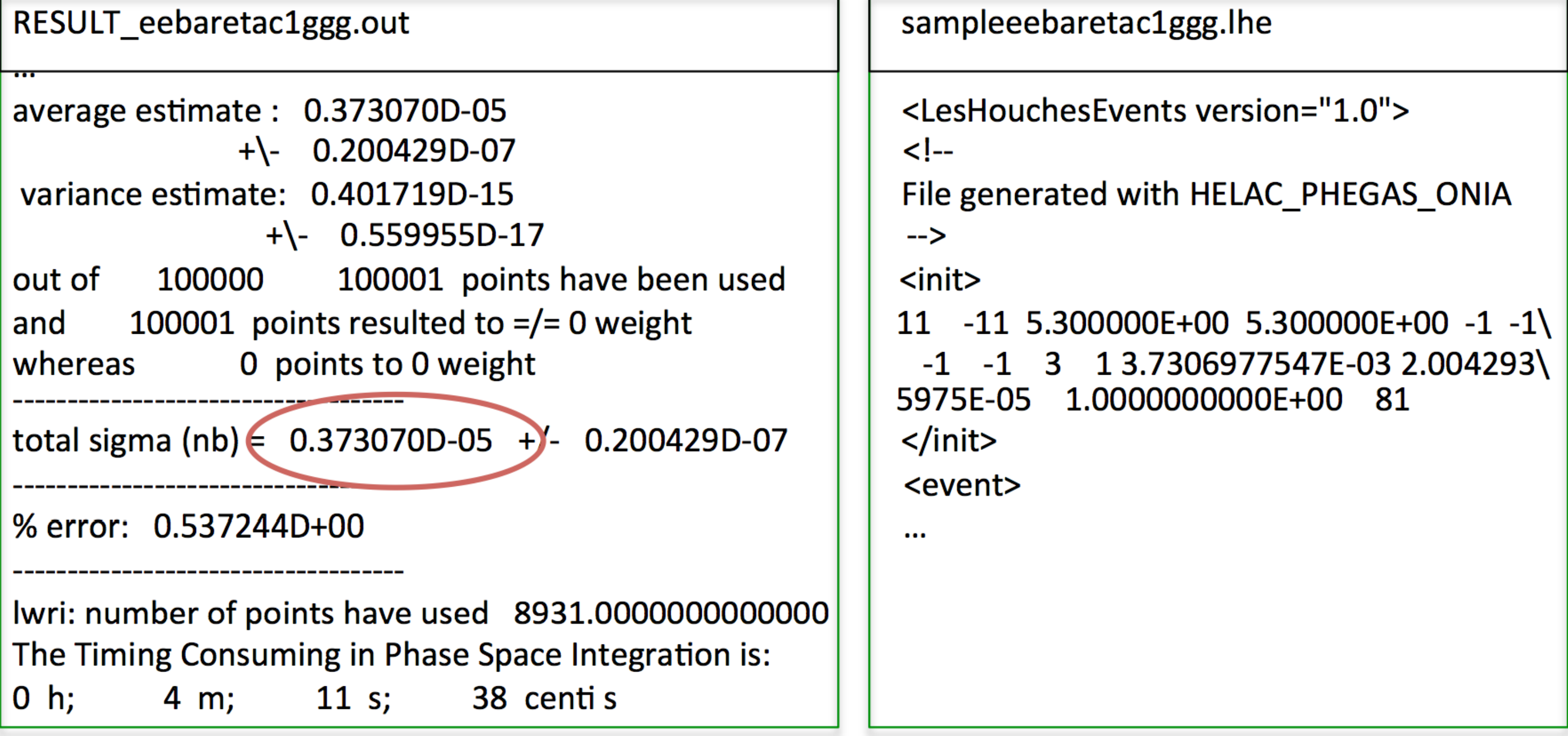}
\caption{\label{fig:output}Illustration of output files for $e^-e^+\to\eta_c+ggg$ with events generated by \mtt{PHEGAS}.}
\end{figure}

The file {\tt user.inp} is left for the user to specify the
parameters in {\tt default.inp} if he/she does not want to use the
default values given in {\tt default.inp}\footnote{We suggest the
user do not change the content in the file {\tt default.inp} unless
he/she really knows what he/she is doing.}. The main parameters are:
\begin{itemize}
\item[(1)]{\tt colpar} represents the type of colliding
particles,i.e. 1 for $pp$,2 for $p\bar{p}$ and 3 for $e^+e^-$.
\item[(2)]{\tt energy} is the center-of-mass energy $\sqrt{s}$ in
unit of GeV.
\item[(3)]{\tt gener} specifies the Monte Carlo generator,i.e. 0 for
\mtt{PHEGAS}, 1 for \mtt{RAMBO}, 2 for \mtt{DURHAM} and 3 for
\mtt{VEGAS},-1 for one phase space point calculation.
\item[(4)]{\tt ranhel} is a parameter to determine whether the program uses the
Monte Carlo sampling over the helicity configurations. In specific,
if {\tt ranhel}$=0$, it does the explicit helicity summation, while if {\tt
ranhel}$>0$, it does the Monte Carlo sampling. If {\tt ranhel}$=1$,
the program uses Monte Carlo sampling over the helicities of the
"elementary" particles in the standard-model and summing over
helicities of quarkonia, while if {\tt ranhel}$=2$ it also performs
Monte Carlo sampling over $\epsilon^{\lambda}_l$ for the P-wave
states, and {\tt ranhel}$=3$ means it does Monte Carlo sampling over
all polarization vectors of heavy quarkonia (of course also over
helicities of the "elementary" particles in the standard-model).
\item[(5)]The value of {\tt qcd} determines the amplitudes should be
calculated in which theory,i.e. 0 for only electroweak, 1 for
electroweak and QCD, 2 for only QCD, 3 for only QED and 4 for QCD
and QED.
\item[(6)] {\tt alphasrun} is a parameter to determine whether the
strong coupling constant $\alpha_S$ should be running(1) or not(0).
\item[(7)] Flags like {\tt gauge}, {\tt ihiggs} and {\tt widsch}
determine the gauge($0=\rm{Feynman~gauge},1=\rm{unitary~gauge}$),
whether inclusion Higgs(1) or not(0) and using the fixed(0) or
complex(1) scheme for the widths of $W^{\pm}$ and $Z$ bosons.
\item[(8)]{\tt nmc} is the number of the Monte Carlo iterations.
\item[(9)]{\tt pdf} is the PDF set number proposed in
LHAPDF\cite{Whalley:2005nh}. Entering $0$ means no PDF is
convoluted.
\item[(10)]{\tt ptdisQ} is a flag whether the $p_T$ distribution of
the first final quarkonium are calculated(T) or just total cross
section(F). If {\tt ptdisQ} is $\rm{T}$, one should also specify
which $p_T$ value ({\tt Pt1}) should be calculated.
\item[(11)]{\tt Scale} specifies which renormalization (and PDF
factorization) scale should be used. It is explained in the comment
line of {\tt default.inp}. If the user chooses the fixed-value
scheme, he/she should also supply the value of the scale ({\tt
FScaleValue}).
\item[(12)]{\tt exp3pjQ} is a flag whether summing over(F)
${\bigl.^{3}\hspace{-1mm}P^{[1/8]}_J},J=0,1,2$ or not(T).
\item[(13)]{\tt modes} determines whether the calculated result is
the polarized one(1) or not(0). If it is $1$, the user should also
supply the values of {\tt SDME1} and {\tt SDME2} to let the program
know which spin density matrix element to calculate. Meanwhile, the
value of {\tt LSJ} represents which "spin" in quarkonium should be
specified. The user should also specify the polarization frame ({\tt
PolarFrame}).
\item[(14)] The parameters of the physical cuts in calculating the
cross sections should also be input by the user if he/she wishes
to use his/her values.
\item[(15)] The long distance matrix elements are also supplied in
{\tt default.inp}. The user can supply his/her values in {\tt
user.inp} with the same format in {\tt default.inp}. The conventions
are explained in the comment lines of {\tt default.inp}.
\end{itemize}
All other parameters are listed in {\tt default.inp}. The user can
fix his/her values in {\tt user.inp} following the format in {\tt
default.inp}. Finally, the user just run the program and obtain the
result files.In Fig.(\ref{fig:output}), we give an illustration of the output files for $e^-e^+\to\eta_c+ggg$,
i.e. {\tt RESULT\textunderscore{}eebaretac1ggg.out} and {\tt sampleeebaretac1ggg.lhe}\footnote{Note that Les Houches Event files can only be generated via {\mtt PHEGAS} in {\mtt HELAC-Onia} now.}. At the end of the first output file, the total cross section is shown in the circle as well as the numerical error, while in the second file is just the event information of the considered process.

\bc
\ec
\section{Summary and outlooks}

The exploitation of the fundamental law in  the nature is the main
aim of the running of the LHC. Heavy quarkonium, which is one type
of the simplest hadrons, provides an ideal laboratory to test and
understand QCD. The discrepancies between the experimental
measurements and the theoretical predictions imply that we still
do not understand the production mechanism of the heavy quarkonium.
Moreover, the quarkonia like $\jpsi$ and $\Upsilon$ production at
the LHC will not only be taken as calibration tools but also be very
useful for TeV physics and even new physics. Hence, it is mandatory
to improve the reliability of Monte Carlo simulation. From the
present theoretical studies, the radiative corrections are very
indispensable even at qualitative level to the quarkonium
production.

Unlike the case in parton-level, the automatic tools for calculating
quarkonium helicity amplitudes are still rare on the market. In this
presentation, we have achieved the first step to the development of
an automatic Monte Carlo generator for heavy quarkonium. Our program
is an extension of the present published
\mtt{HELAC}~\cite{Kanaki:2000ey,Papadopoulos:2006mh,Cafarella:2007pc},
which is based on an off-shell recursive algorithm or
Dyson-Schwinger equations. We dubbed it as \mtt{HELAC-Onia}. It
provides an automatic computation tool for heavy quarkonium helicity
amplitudes in the standard model with high efficiency. We have also
shown the applications of our tool to the various aspects of the
heavy quarkonium production from $pp$,$p\bar p$ and $e^+e^-$
collisions.

The following steps are to realize the automation of the
next-to-leading order quarkonium helicity amplitudes computations.
With such a code, one can perform the analysis of the heavy
quarkonium production at a full next-leading order level, which is
much more reliable and useful especially at the LHC.

\vfill


%





\noindent{\Large \bf Acknowledgements} \\[24pt]
This work was supported in part by the National Natural Science
Foundation of China (Nos.11021092,and 11075002), and the Ministry of
Science and Technology of China (No.2009CB825200). First of all,we
are grateful to Costas Papadopoulos, Malgorzata Worek and the HELAC team to allow us
using the acronym and encourage us to upload the program on the
HELAC web-page. We would also like to thank Prof. Kuang-Ta Chao for
providing important supports on this project, Qiang Li for useful
discussion on matrix element and parton shower matching and Pierre Artoisenet for using \mtt{MADONIA}.

\vspace*{2cm}


\providecommand{\href}[2]{#2}\begingroup\raggedright\endgroup

\end{document}